\documentclass[useAMS,usenatbib]{mn2e}
\usepackage{latexsym,graphicx,natbib,amssymb}

\title[Evidence for dark matter in UCDs]
 {High mass-to-light ratios of UCDs - Evidence for dark matter ?}
      
\author[H. Baumgardt and S. Mieske]
{H. Baumgardt$^{1}$ and S. Mieske$^2$\\
 $^1$ Argelander-Institut f\"ur Astronomie, Universit\"at Bonn, Auf dem H\"ugel 71, 53121 Bonn, Germany \\
 $^2$ European Southern Observatory, Karl-Schwarzschild-Strasse 2, 85748 Garching bei M\"unchen, Germany
}

\date{Accepted ????. Received ?????; in original form ?????}

\pagerange{\pageref{firstpage}--\pageref{lastpage}} \pubyear{2008}

\begin{document}   

\maketitle

\label{firstpage}

\begin{abstract}
  Ultra-compact dwarf galaxies (UCDs) are stellar systems with masses
  of around $10^7$ to $10^8$ M$_\odot$ and half mass radii of 10-100 pc. They
  have some properties in common with massive globular clusters,
  however dynamical mass estimates have shown that UCDs have
  mass-to-light ratios which are on average about twice as large than
  those of globular clusters at comparable metallicity, and tend to be larger
  than what one would expect for old stellar systems composed out of
  stars with standard mass functions.
  
  One possible explanation for elevated high mass-to-light ratios in UCDs
  is the existence of a substantial amount of dark matter, which
  could have ended up in UCDs if they are the remnant nuclei of
  tidally stripped dwarf galaxies, and dark matter was dragged into
  these nuclei prior to tidal stripping through e.g. adiabatic gas
  infall.  Tidal stripping of dwarf galaxies has also been suggested
  as the origin of several massive globular clusters like Omega Cen,
  in which case one should expect that globular clusters also form
  with substantial amounts of dark matter in them.
  
  In this paper, we present collisional N-body simulations which study
  the co-evolution of a system composed out of stars and dark matter.
  We find that the dark matter gets removed from the central regions of such
  systems due to dynamical friction and mass segregation of stars. 
  The friction timescale is significantly shorter than a Hubble time
  for typical globular clusters, while most UCDs have friction times much 
  longer than a Hubble time. Therefore, a significant dark matter 
  fraction remains within the half-mass radius of present-day UCDs, making
  dark matter a viable explanation for the elevated M/L ratios of UCDs. If 
  at least some globular clusters formed in a way similar to UCDs, we predict 
  a substantial amount of dark matter in their outer parts.
\end{abstract}


\section{Introduction}
\label{sec:intro}

Ultra-compact dwarf galaxies (UCDs) were discovered in the late 1990s
in spectroscopic surveys of the Fornax galaxy cluster \citep{hetal99,
detal00} and have since then been found in other nearby galaxy
clusters as well
\citep{hetal05,metal05,metal07,jetal06,fetal07,retal07}. They are
bright ($-11 < M_V < -13.5$) and compact ($7 < r_h < 100$ pc) stellar
systems which have ages of at least several Gyr and possibly up to 10
Gyr \citep{metal06,eetal07}.  The masses and sizes of UCDs are larger
than those of Galactic globular clusters, but similar to those of
nuclei in dwarf elliptical galaxies (Drinkwater et al. 2003, Bekki et
al. 2003).

One of the most remarkable properties of UCDs is that their dynamical
mass-to-light ratios are on average about twice as large than those of
globular clusters of comparable metallicity, and also tend to be
larger than what one would expect based on simple stellar evolution
models that assume a standard stellar initial mass function, like e.g.
\citet{k01} \citep{hetal05,dhk08,metal08}. If not due to a failure of
stellar evolution models, this points either to unusual stellar mass
functions \citep{mk08, dbk08} or possibly to the presence of a
significant amount of dark matter in UCDs. We note that most methods
used to determine M/L ratios for UCDs rely on the assumptions that mass follows
light and isotropic velocity dispersions. How well these assumptions are fulfilled is currently
not known. Also due to the large
distances of UCDs, only integrated velocity dispersions can be obtained, which
in most cases are intermediate between the central and the global velocity dispersions. 
In order to determine the mass-to-light ratio, the mass modeling has to take 
the density profiles of the UCDs as well as the effects of seeing and a finite slit size
into account, as done for example in \citet{hetal07}.

\begin{figure}
\begin{center}
\includegraphics[width=8.5cm]{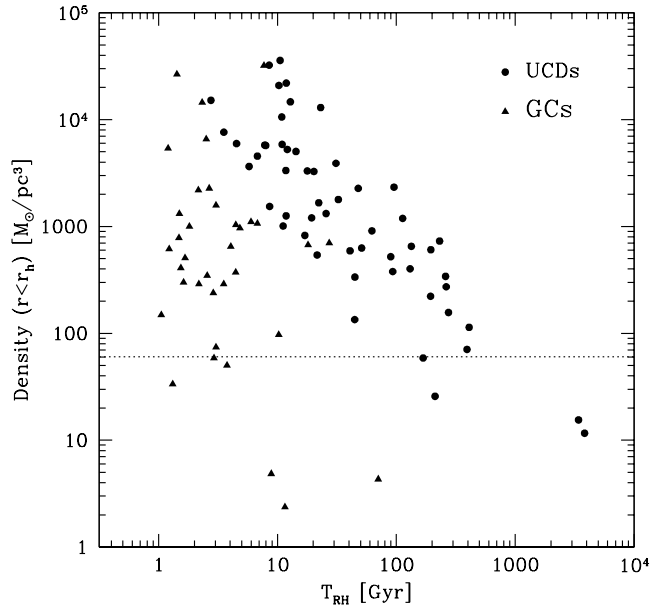}
\end{center}
\caption{The mean mass density within the half-mass radius of the
joint sample of GCs and UCDs from Fig.~\ref{fig:mtol2} is plotted
vs. their relaxation time \citep{metal08}. The dotted line indicates
the approximate central ($r\lesssim10$pc) dark matter densities
expected for cuspy dwarf galaxy CDM halos \citep{giletal07}. }
\label{fig:density}
\end{figure}

Several formation scenarios have been discussed for UCDs, like e.g.\
that UCDs are simply massive globular clusters and form in the same
way \citep{hetal99,eetal07,fetal08}, that they are the nuclei of tidally
stripped, originally much more extended galaxies
\citep{bcd01,bcd03,tde08,getal08}, or that they are merged globular
clusters \citep{ol00,fk02}. \citet{getal08} have shown that funneling
of dark matter to the central region of a disk galaxy, due to
gas-infall, can significantly increase the M/L ratios in the nuclear
region, and hence may explain the elevated M/L ratios of UCDs,
provided that UCDs formed by tidal stripping. Indeed, it has been
suggested that also GCs may have originated as centers of individual
primordial dark matter halos (e.g. Carraro \& Lia 2000, Lee et al. 2007, Bekki et al.
2007). If dark matter funneling is an efficient mechanism
\citep{getal08}, one may therefore expect both UCDs and GCs to be
formed with a significant fraction of dark matter. It is important to
note that such an increase of dark matter density by some kind of
funneling mechanism is necessary to explain a significant amount of
dark matter in UCDs or GCs, since their present-day stellar (and hence
implied dark matter) densities are up to 2-3 orders of magnitude
higher than expected for cuspy dark matter halos of dwarf galaxy mass
\citep{giletal07}. This is shown in Fig.~\ref{fig:density}.

In this paper, we start from the working hypothesis that both GCs and
UCDs are formed with the same non-zero dark-to-stellar-mass-fraction.
We then investigate how the dynamical co-evolution of dark matter and
stars changes the observed dark matter fraction as a function of time.
We assess whether the observed rise of M/L ratios from the
regime of GCs to that of UCDs can be explained by our working
hypothesis and the subsequent dynamical evolution.

\section{The models}
\label{sec:Nbody}

In our simulations, we assume that stars and dark matter particles follow the same density
distribution initially, which was given by a \citet{p11} model.
Determinations of mass-to-light ratios of globular clusters or UCDs
rely mainly on stars located inside the half-mass radius
\citep{hetal07,metal08} or even closer within \citep{mcetal05}. Tidal
effects are therefore not likely to have a strong influence on
determined mass-to-light ratios. Also, due to their high mass and
corresponding large dissolution times, tidal interactions probably play only a minor role
for UCDs. In our simulations we therefore neglect the influence of an external tidal field. 

We assume
that the stars initially follow a \citet{k01} mass function with lower
and upper mass limits of 0.1 and 100 M$_\odot$.  Stellar evolution
changes the mass function of stars, however most of this change
happens within the first $10^9$ yrs (see e.g. the grid of models by
\citet{bm03}), i.e. on a timescale short against the lifetime of
globular clusters or UCDs. We therefore also neglect stellar evolution
and immediately transform stars to the assumed age of GCs and UCDs, T=12 Gyrs.  
For the transformation, we assume that stars
with mass larger than 25 M$_\odot$ form black holes and assume that
the black hole mass is 10\% of the mass of the initial star.  This
way, black hole masses in our models are compatible with observed
masses for stellar mass black holes (e.g.\ \citet{c06}). Stars with
masses between 8 M$_\odot$ and 25 M$_\odot$ are assumed to form
neutron stars with a mass of $m=1.3$ M$_\odot$ \citep{tc99}. Stars
between 0.8 M$_\odot$ and 8 M$_\odot$ are assumed to form white dwarfs
due to stellar evolution. The masses of the white dwarfs are obtained
from \citet{k08}, who found, based on observations of white dwarfs in
star clusters, the following relation between the initial and final
mass of white dwarfs: $m_{wd} = 0.109 m + 0.394 M_\odot$. Stars less
massive than 0.8 M$_\odot$ are still on the main sequence and we
assume that they have not yet lost any mass.  The following table
summarises our initial-to-final mass relation:
\begin{equation}
m_{rem} = 
   \left\{ \begin{array}{l} 
       m, \;\;\;\; m < 0.8 M_\odot \\[+0.2cm] 

       0.109 \frac{m}{M_\odot} + 0.394 , \;\;\;\; 0.8 M_\odot < m < 8 M_\odot \\[+0.2cm]

       1.35, \;\;\;\; 8 M_\odot < m < 25 M_\odot \\[+0.2cm]

        0.1 m, \;\;\;\; 25 M_\odot < m  \end{array} \right. 
\end{equation}

Neutron stars and probably also black holes receive kicks at the time
of their birth due to asymmetric supernova explosions. The size of
these kicks is a few hundred km/sec \citep{ll94}, which is large
enough that most will be lost from globular clusters or UCDs
\citep{p02}. We therefore assume only a small neutron star and black hole
retention fraction of 20\% in our simulations.  With these
assumptions, the mean mass of stars in our models is 0.344 M$_\odot$.

The dark matter is also modeled as point mass particles. In our reference
simulation we assume a mass of 0.03 M$_\odot$ for the dark matter
particles, but we also make simulations with masses of 0.15 M$_\odot$
and 0.01 M$_\odot$ to study the influence of the adopted particle mass
on our results. Theoretically, the dynamical friction of stars should
not depend on the mass of the dark matter particles as long as the
mass ratio between stars and dark matter is high enough
\citep{bt87}. Also, with the adopted masses, the self-interaction of
the dark matter particles is still unimportant over the timescales
studied here. We will further investigate the influence of the mass of
the dark matter particles in sec. \ref{sec:variations}.

All runs are performed with the collisional $N$-body code NBODY4
\citep{a99} on the GRAPE6 computers \citep{mfkn03} of Bonn
University and the results are expressed in $N$-body units \citep{hh02},
in which the constant of gravity and total cluster mass are equal to 1
and the total potential energy is equal to -0.5. Table~1 gives an
overview of the runs performed.

\section{Results}
\label{sec:results}

\subsection{Mass segregation timescale: analytical estimate}

In the following we derive an analytical estimate of the mass segregation time
scale in compact stellar systems. 

Massive stars will segregate against dark matter particles and lighter
stars as a result of dynamical friction and energy
equipartition. Since the masses of stars are much higher than the mass
of the dark matter particles, the frictional drag on the stars is
given by (see \citet{bt87} Eq.\,7-18):
\begin{equation}
 \frac{d\vec{v}}{dt} = -\frac{-4 \pi \ln{\Lambda} G^2 \rho(r) m}{v^3} 
\left[ {\rm erf}(X) - \frac{2 X}{\sqrt{\pi}} e^{-X^2} \right] \vec{v}
 \label{dynf}
\end{equation}
where $\rho(r)$ is the background density of dark matter and stars,
$m$ the mass of an inspiraling star, $\ln{\Lambda}$ the Coulomb
logarithm, and $X=\vec{v}/(\sqrt{2}\sigma)$ is the ratio between the
velocity of a star and the (1D) stellar velocity dispersion
$\sigma$. If we assume $v \approx \sigma$, it follows that
$X=1/\sqrt{2}$. Setting $\ln{\Lambda}=12$ for globular clusters
(\citet{bt87} Tab.\ 7-1), the eq.\ \ref{dynf} can be rewritten as:
\begin{equation}
 \frac{dv}{dt} = - 29.96 \frac{G^2 \rho(r) m}{v^2}\,.
\end{equation}
The resulting energy change is $\frac{dE}{dt}= \frac{d}{dt}
(\frac{1}{2} m v^2) = m v \frac{dv}{dt}$. For a distribution of stars
in virial equilibrium, an energy change $dE$ corresponds to a change
in potential energy $m d\Phi =2 dE$. Hence
\begin{equation}
 \frac{d\Phi}{dt} = - 59.93 \frac{G^2 \rho(r) m}{v}\,.
\end{equation}
For a Plummer model, the density $\rho$, circular velocity $v_c$ and
specific potential $\Phi$ at point $r$ are given by:
\begin{eqnarray}
\nonumber \rho(r) & = & \frac{3 M_{Tot} a^2}{4 \pi}  \left( a^2 + r^2 \right)^{-5/2} \\
 \Phi(r) & = & - \frac{G M_{Tot}}{\left( a^2 + r^2 \right)^{1/2}} \\ 
\nonumber v_c(r) & = & \frac{\sqrt{G M_{Tot} r^2}}{\left( a^2 + r^2 \right)^{3/4}} 
\end{eqnarray}
where $M_{Tot}$ is the total cluster mass and $a$ is the scale radius of the Plummer model. With these equations, 
the above relation can be rewritten as:
\begin{equation}
\frac{d r}{dt} = -14.31 \frac{\sqrt{G} a^2 m}{\sqrt{M_{Tot}} r^2 \left( a^2 + r^2 \right)^{1/4}}
\end{equation}
For an order of magnitude estimate of the inspiral time scale, one can approximate $\left( a^2 + r^2 \right)^{1/4} \approx a^{1/2}$.
One can then solve the above relation and obtains as time which a star starting at radius $R_0$ needs to reach the centre:
\begin{equation}
\nonumber T_{\rm Fric} = 0.023 \frac{\sqrt{M_{Tot}}}{\sqrt{G} a^{3/2} m} R_0^3 
\end{equation}
For a Plummer model, $a=0.766 R_H$, so the inspiral time for stars near the half-mass radius is given by:
\begin{equation}
\nonumber T_{\rm Fric} = 0.035 \frac{\sqrt{M_{Tot}} R_H^{3/2}}{\sqrt{G} m} 
\label{tfric}  
\end{equation}
For a globular cluster or UCD we thus obtain:
\begin{equation}
\nonumber T_{\rm Fric} = 5.86 \left(\frac{M_{Tot}}{10^6 M_\odot}\right)^{1/2}  \left(\frac{R_H}{5 pc}\right)^{3/2}
 \left(\frac{m}{M_\odot} \right)^{-1} \rm{Gyr}
\label{tfricgc}
\end{equation}
The resulting dynamical friction time agrees to within 20\% with what
Binney \& Tremaine found for the inspiral time of an isothermal sphere
(their eq. 7-26).  Eq.\ \ref{tfricgc} predicts a dynamical friction
time scale of 4-5 Gyr for a typical GC (3$\times 10^5$M$_{\odot}$,
$r_h=3$pc), and about 400 Gyr for a typical UCD ($10^7$M$_{\odot}$,
$r_h=20$pc). That is, after a Hubble time most of the dark matter in
globular clusters should have been pushed out of the centre, while in
UCDs the inspiral of stars should be far from complete and a
significant fraction of DM should still reside in their centres,
leading to high mass-to-light ratios.

\subsection{$N$-body results}
\label{sec:res-nbody}

\begin{figure}
\begin{center}
\includegraphics[width=8.5cm]{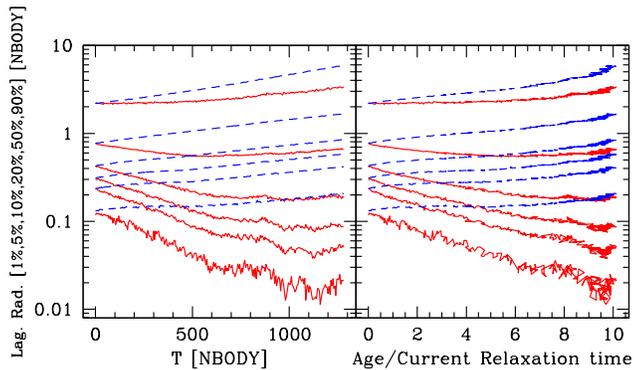}
\end{center}
\caption{Evolution of Lagrangian radii, i.e. radii which contain a certain fraction of the total mass, of stars (red solid lines) and 
dark matter particles (blue dashed lines) in
the first run from Table~1. The left panel depicts the evolution as a function of $N$-body time, the right panel as a function of
current relaxation time, where the relaxation time is calculated based on the distribution of stars only. Inside the half-mass radius
of the cluster, less than 20\% of the total mass is made up out of dark matter after the clusters are ten apparent
relaxation times old.}
\label{fig:lagrad}
\end{figure}

Fig.\ \ref{fig:lagrad} shows the evolution of Lagrangian (lag.) radii,
i.e. radii which contain a certain fraction of the total mass,
of stars and dark matter particles in our first simulation, which had
a dark matter particle mass of $m=0.03$ M$_\odot$ and an equal amount
of mass in stars and dark matter. The effect of dynamical friction and
mass segregation is clearly visible since the lag. radii of dark
matter particles increase with time while those of the stars shrink.
In $N$-body units, the total cluster mass and constant of gravity are
both unity. With a mean stellar mass of $m=6.0 \cdot 10^{-5}$, we
predict a dynamical friction timescale of $T_{\rm Fric} = 391$ in
$N$-body units according to eq.\ \ref{tfric}. It can be seen that by
this time the cluster center is indeed nearly free of dark matter:
inside the lag. radius of 10\% of the stars, only 1\% of the dark matter
particles are located. At the end of the simulation, the 10\% lag.
radius of the dark matter particles is almost equal to the half-mass
radius of the stars, i.e.\ only 20\% of the cluster mass is still made
up of dark matter inside the (visible) half-mass radius of the
cluster. The core of the cluster is even stronger depleted and is
nearly free of dark matter by the end of the simulations.
It would therefore be difficult to detect
the remaining dark matter by its effect on stellar velocities if
mainly stars from the cluster center or inside the half-mass radius
are used to determine the line-of-sight velocity dispersion.

The right panel of Fig.\ \ref{fig:lagrad} depicts the evolution of
lag. radii as a function of the ratio of cluster age to the actual relaxation
time. In order to allow for a better comparison with observations, the
relaxation time is calculated from the stellar component according to \citep{s87}:
\begin{equation}
 T_{RH} =  0.138 \frac{\sqrt{M_*} R_{H *}^{3/2}}{\sqrt{G} m_* \ln{\gamma N_*}}
\label{eq:trels}
\end{equation}
where $M_*$ is the total stellar mass of the cluster, $R_{H *}$
the half-mass radius of the stellar distribution, $m_*$ and $N_*$
are the mass and number of stars and $\gamma$ a constant in the Coulomb logarithm
which is taken to be $\gamma=0.11$ \citep{gh94}.
Eq.\ \ref{eq:trels} would be the relaxation time
inferred by an observer who can only determine the stellar
distribution and does not know about the dark matter. It has
the same dependence on cluster mass and radius as the friction timescale and can
therefore also be used to judge the dynamical state of a cluster.

The right panel of Fig.\ \ref{fig:lagrad} shows
that once a cluster is two to three apparent relaxation times old, the
centre is free of dark matter and by the time the cluster has become
ten relaxation times old, there is little dark matter left inside the
half-mass radius.  Since most globular clusters have relaxation times
of only a few Gyr, their mass-to-light ratios should be within
20\% of those of pure stellar populations if mainly stars inside the
clusters half-mass radius are used to determine the velocity
dispersion. Since this is within the uncertainty of measured
mass-to-light ratios and current stellar population models, such a
small dark matter cannot be detected kinematically in globular
clusters.  UCDs on the other hand have relaxation times significantly
larger than a Hubble time and should therefore still have large
mass-to-light ratios if they formed as a mix of dark matter and stars.

\begin{figure}
\begin{center}
\includegraphics[width=8.5cm]{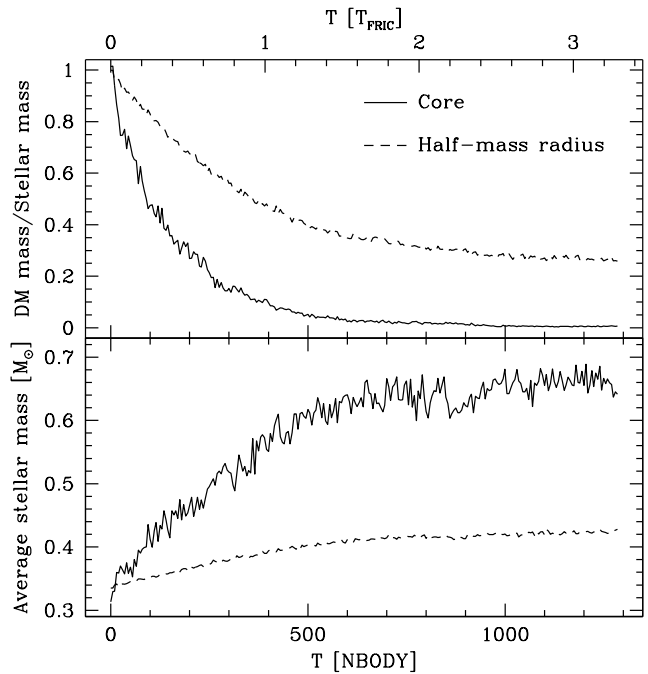}
\end{center}
\caption{Dark matter fraction (upper panel) and average mass of stars (bottom panel) as a function of time.
Core values are shown by solid lines, values inside the half-mass radius by dashed lines. The core radius is assumed
to be equal to the 5\% lag. radius of the stellar distribution. The dark matter is depleted from the centre within
1 to 2 friction times. At the same time, heavy mass stars segregate against the light mass stars and the average
mass of stars increases in the centre. Clusters where the centers are depleted of dark matter should therefore also be 
mass segregated in their centre.}
\label{fig:mcore}
\end{figure}

This is confirmed by the upper panel of Fig.\ \ref{fig:mcore}, which
depicts the dark matter fraction inside the cluster core (assumed to
be the region inside the 5\% lag. radius of the stars) and inside the
half-mass radius of the cluster stars. It can be seen that after about
1.5 to 2 friction times, the cluster core is almost completely
free of dark matter.  Within the half-mass radius, only about 30\% of
the initial dark matter amount remains after this time. The lower
panel of Fig.\ \ref{fig:mcore} depicts the evolution of the average
mass of stars in the core and inside the half-mass radius.  At both
radii, the average mass of stars is increasing since, while stars
segregate against the dark matter particles, heavy-mass stars also
segregate against the lighter ones. After about 2 dynamical friction
timescales, the mass of stars has reached a near constant value of
about 0.65 M$_\odot$, which is significantly higher than the average
mass of stars.  Clusters which have expelled dark matter out of their
centres should therefore also be mass segregated.

\subsection{Comparison with observations}
\label{sec:comp}

\begin{figure}
\begin{center}
\includegraphics[width=8.5cm]{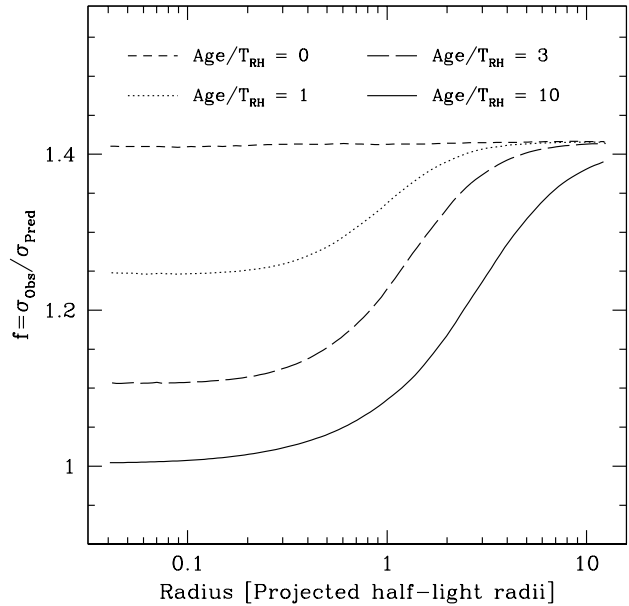}
\end{center}
\caption{Evolution of the ratio $f$ of observed velocity dispersion to predicted
one based on the stellar distribution alone for 4 different ratios of cluster age to 
apparent relaxation time calculated according to eq.\ \ref{eq:trels}. Initially, dark matter
and stars follow the same distribution and there is an equal amount of mass in dark 
matter and stars, so $\sigma_{Obs} = \sqrt{2} \sigma_{Pred}$ independent of radius.
As the dark matter is expelled from the centre, stellar velocities in the inner parts are increasingly determined 
by the stars alone and $f$ drops towards unity in the cluster centre. Globular clusters 
should therefore have central mass-to-light ratios in agreement with stellar population models.}
\label{fig:velfac}
\end{figure}

Fig.\ \ref{fig:velfac} shows the effect which the decreasing dark matter fraction in the center has on the projected
velocity dispersion of stars. In order to determine this effect, we first calculate the velocity 
dispersion profile $\sigma_{Obs}(r)$ of bright stars with masses in the range $0.6 < m < 0.9$ M$_\odot$ as a function of 
projected radius.
We restrict ourselves to this mass range since in a globular cluster or UCD, these would be the stars which dominate
the cluster light. After determing the velocity dispersion profile of bright stars,
we calculate the expected velocity dispersion profile based on the stellar density distribution
according to (Binney \& Tremaine 1987, eq. 4-54):
\begin{equation}
 \sigma^2(r) = - \frac{1}{\rho(r)} \int \rho(r') \left. \frac{d\Phi}{dr}\right|_{r=r'} dr'
\label{eq:jeans}
\end{equation}
where $\rho(r)$ is the (3D) density distribution of bright stars and $\Phi(r)$ is the potential coming
from the stars alone. Eq.\ \ref{eq:jeans} assumes a spherical cluster potential and an isotropic velocity
dispersion of stars. After projecting $\sigma(r)$ we can calculate the correction factor $f$ needed so that the 
predicted velocity dispersion matches the true velocity dispersion of the clusters in the $N$-body 
simulations, i.e. $f(r)=\sigma_{Obs}(r)/\sigma_{Pred}(r)$. The resulting correction factor is plotted in Fig.\ 
\ref{fig:velfac} for the first run from Table~1. Initially, dark matter and stars follow the same density distribution, 
so the velocity dispersion is a factor $f(r)=\sqrt{(M_{DM}+M_*)/M_*}=1.41$ higher than predicted by eq.\ \ref{eq:jeans}.
As the cluster evolves, dark matter is removed from the center, so the velocities of stars in the center
are determined more and more by the stars alone and $f$ approaches unity. After 3 relaxation times, the
central velocity dispersion is only 10\% higher than what one would expect based on the stars alone and after
10 relaxation times the difference is less than 1\%. Most globular clusters should therefore have central M/L ratios 
which are close to those predicted by stellar population models. Beyond 10 half-mass radii, $f$ remains close to the
initial value even after 10 relaxation times. As long as dark matter is not removed by tidal effects \citep{ms05}, it should 
therefore be detectable in globular clusters through the observation of stellar velocities in the outer cluster
parts.

We finally discuss the influence of the dark matter on the global mass-to-light ratios.
In order to compare our simulations with observed clusters, we again
calculate true and expected velocity dispersions of stars with masses in the range $0.6 < m < 0.9$ M$_\odot$.
Since mass-to-light ratios of UCDs are determined from stellar velocities
covering a significant fraction of the cluster area (see e.g.\ discussion in
\citet{hetal07}) and measured mass-to-light ratios of globular clusters are
based mainly on stars in the inner cluster parts \citep{mcetal05}, we determine 
global velocity dispersions in the simulations for all stars located inside the projected half-light
radius. The resulting mass-to-light ratios of our model clusters are then given by
\begin{equation}
 M/L = f^2 M/L|_*
\end{equation}
where $M/L|_*$ is the mass-to-light ratio which a pure stellar population would have and $f$ is again
$f=\sigma_{Obs}/\sigma_{Pred}$.

\begin{figure}
\begin{center}
\includegraphics[width=8.5cm]{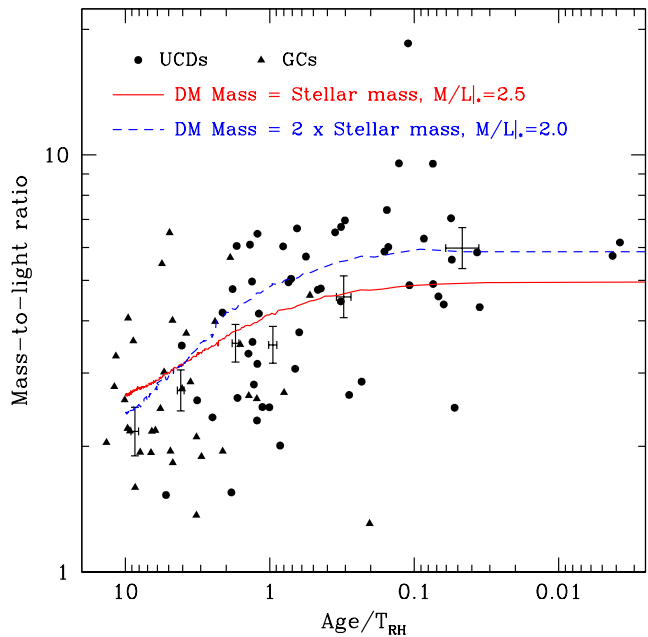}
\end{center}
\caption{Mass-to-light ratios of UCDs (circles) and globular clusters (triangles) as a function of their age
divided by their relaxation time. There is a clear trend towards lower M/L values for dynamically more evolved
systems. The red solid and blue dashed curves show predicted M/L values for two of our runs calculated from
the velocity dispersion of bright stars inside the clusters half-mass radius, assuming stellar mass-to-light
ratios of $M/L|_*=2.0$ and $M/L|_*=2.5$ for the two runs. It can be seen that the
resulting theoretical curves provide a good fit to the combined globular cluster/UCD sample.}
\label{fig:mtol2}
\end{figure}

Fig.\ \ref{fig:mtol2} depicts the evolution of $M/L$ with cluster age for runs 1 and 2 of
Table~1, and compares it with observed M/L ratios of UCDs
and GCs from \citet{metal08}. Note that the literature M/L estimates are
normalised to the same (solar) metallicity, to allow direct
intercomparison. Time is again expressed in terms of age divided by
the relaxation time as determined from the stars alone.

The observed normalised M/L ratios show a clear trend in the sense that dynamically more evolved
systems have on average lower M/L values. The mass-to-light ratios in our simulations 
also decrease as the dynamic age increases, since, as the dark matter is depleted from the cluster
centers, the velocity dispersion is determined more and more by the stars alone, so M/L approaches
$M/L|_*$. Depending on whether a stellar $M/L|_*$ of 2.5 or 2.0 is assumed, a run with
a primordial dark matter content equal to or twice as high as the
stellar mass provides an acceptable fit to the data, making dark matter a
viable alternative to explain the elevated mass-to-light ratios of
UCDs. We note that if the observed decrease of M/L with dynamical age is due to the dynamical depletion of 
non-luminous particles from the cluster centers, the dark matter particles have to 
be of lower mass than the stars, ruling out e.g. a central concentration of black holes 
in UCDs as the explanation for their high M/L ratios.

The stellar mass-to-light ratios we have to
assume in order to fit the data for GCs are marginally lower than
predicted by simple stellar population models for 12 Gyr old,
solar-metallicity star clusters. For example, \ the \citet{bc03}
models predict an M/L of 2.5 for a 8-9 Gyr old stellar population. The
difference to a 12 Gyr old population (M/L=3.5) is still within
individual error bars for the literature estimates, and there is also
some uncertainty in the underlying stellar mass functions and stellar
population models. Nevertheless, we note that the slightly too low M/L
ratios may also be interpreted as signs of preferential loss of low-mass stars in the
Galactic tidal field \citep{bm03, kr08}. If this was the case, then less
dark matter would be needed to explain the elevated mass-to-light
ratios of UCDs, implying a DM mass of $\sim$50-80\% of the stellar
mass. However, it is unclear whether the actual dissolution times of
the GCs with available M/L measurements are short enough to have experienced
significant evaporation \citep{metal08}. A case-by-case analysis for
Galactic GCs will be necessary to assess this, based on measured
absolute proper motions and orbital parameters \citep{aetal06}.
\begin{figure}
\begin{center}
\includegraphics[width=8.5cm]{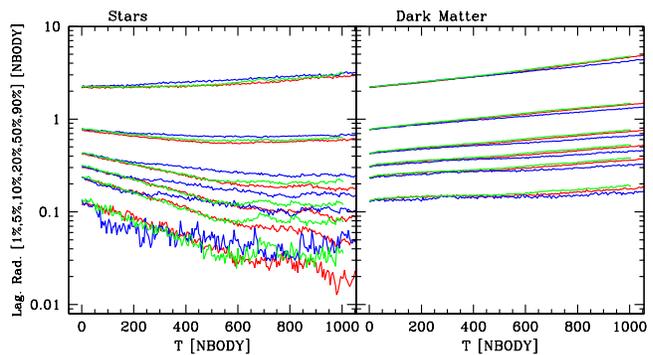}
\end{center}
\caption{Evolution of Lagrangian radii of stars (left panel) and dark
matter particles (right panel) in simulations which assume dark matter
particle masses of $m=0.1$ M$_\odot$ (blue lines), $m=0.03$ M$_\odot$
(red lines, the default mass for the simulations presented in
Figs.~\ref{fig:lagrad} to~\ref{fig:mtol2}) and $m=0.015$ M$_\odot$
(green lines). The agreement between the different curves, especially
for light dark matter particles is very good, showing that the adopted
mass $m=0.03$ M$_\odot$ of the dark matter particle should not influence our results.}
\label{fig:dmpar_var}
\end{figure}

\subsection{Scaling issues}
\label{sec:variations}

We finally discuss a possible biasing of our results due to the finite mass of the dark matter
particles. Fig.\ \ref{fig:dmpar_var} depicts the evolution of Lagrangian radii 
in simulations with different dark matter particle masses. All simulations
had an equal amount of mass in stars and dark matter
and the mass of the dark matter particles was set to be $m=0.1$
M$_\odot$ (blue lines), $m=0.03$ M$_\odot$ (red lines) and $m=0.01$
M$_\odot$ (green lines). Since the mass of heavy stars which drive the 
inspiral is in all cases much higher than the mass of the dark matter particles,
eq.\ \ref{tfric} should still apply for the inspiral timescale. In all three simulations,
the mass of individual stars if expressed in $N$-body units was the same, so 
according to eq.\ \ref{tfric}, the inspiral timescale of the stars should be the same in 
the three simulations.

It can be seen that the inspiral of stars and the ejection of dark
matter particles happens in all three clusters in a very similar way.
The agreement is especially good between the two simulations with the
lightest dark matter particles. We therefore conclude that the adopted
mass $m=0.03$ M$_\odot$ for the dark matter particle does not
influence the results presented in Figs.~\ref{fig:lagrad}
to~\ref{fig:mtol2}. Our simulations should therefore give a correct picture of
the dynamical ejection of dark matter from the centers of globular
clusters and UCDs.

\section{Conclusions}
\label{sec:concl}

We have performed collisional $N$-body simulations of the evolution of compact systems
composed out of a mix of stars and dark matter particles. Our simulations show that
dark matter is depleted from the centers of these systems due to dynamical friction and
energy equipartition between stars and dark matter particles. The inspiral time of stars
is short enough that only 20\% of the original dark matter would remain within the
half-mass radius in typical globular clusters. If mainly stars from the inner cluster parts are
used to determine mass-to-light ratios, the resulting increase in the mass-to-light
ratio is within the errors with which mass-to-light ratios are typically determined 
for globular clusters and would therefore be difficult to detect. 

If not tidally stripped, 
dark matter should also reside in the outer parts of globular clusters. For a number of
globular clusters, \citet{setal07} have indeed reported a flattening of the velocity disperion in 
the outer cluster parts. This could however be due to a number of reasons like contamination of the 
sample by background stars or the tidal interaction of a star cluster with the gravitational field of 
the Milky Way \citep{detal98, cetal05}. Detailed simulations would be necessary to exclude 
these possibilities and confirm that the observed flattening is due to a dark matter halo.

UCDs on the other hand have inspiral times 
significantly longer than a Hubble time and therefore still contain most of the dark matter 
in their centers. Dark matter therefore seems a viable explanation for the
elevated M/L ratios of UCDs, provided that UCDs originate from the
centers of dark matter halos and have seen their dark matter content
being increased by dark matter funneling, through e.g. adiabatic gas
infall \citep{getal08}. 

A prediction of our simulations, which can in principle be tested by
observations, is that globular clusters which have expelled the dark
matter from their centers should also be mass segregated. Non-mass
segregated clusters with velocity dispersions and mass-to-light ratios
in agreement with simple stellar population models, would therefore
have formed without significant amounts of dark matter in their
centers.

Also, if dark matter existed in a globular cluster at the time of its
formation, it should still reside in its outer parts, especially if
tidal stripping due to external tidal forces from the host galaxy
\citep{ms05} and relaxation driven internal mass loss was not
important for the cluster evolution.  In this case, the measured
mass-to-light ratio should increase towards the outer cluster parts,
which can in principle be detected with dedicated radial velocity or
proper motion surveys. The future astrometric satellite
{\it GAIA} would be an excellent tool for such a search since it will
provide accurate proper motions for thousands of stars in the halos of
nearby globular clusters. 

\section*{Acknowledgements}

\label{lastpage}

\begin{table*}
\caption[]{Details of the $N$-body models. The second column gives the initial number of stars, the
third column the initial number of dark matter particles. The fourth column gives the mass of a dark
matter particle and the fifth column gives the relative mass fraction in dark matter and in stars. The last columns give 
the fraction of dark matter remaining inside the 5\% lagrangian radius and inside the half-mass radius after 
one, two and ten apparent relaxation times (eq.\ \ref{eq:trels}) have passed.}
\begin{tabular}{rcrccccccccc}
 Nr. & $N_*$ & \multicolumn{1}{c}{$N_{DM}$} & $m_{DM}$ & $M_*:M_{DM}$ & \multicolumn{2}{c}{$f_{DM}|_{T=T_{RH}}$} & 
  \multicolumn{2}{c}{$f_{DM}|_{T=2 T_{RH}}$} & \multicolumn{2}{c}{$f_{DM}|_{T=10 T_{RH}}$} \\
 & & & [$M_\odot$] & & $r<R_{5\%}$ & $r<R_H$ & $r<R_{5\%}$ & $r<R_H$ & $r<R_{5\%}$ & $r<R_H$  \\
 1 & 8224 & 91776 & 0.030 & 1:1 & 0.39 & 0.78 & 0.22 & 0.63 & 0.01 & 0.26 \\ 
 2 & 4285 & 95715 & 0.030 & 1:2 & 0.57 & 1.51 & 0.29 & 1.27 & 0.03 & 0.55 \\
 3 & 8223 & 27528 & 0.100 & 1:1 & 0.38 & 0.81 & 0.20 & 0.70 & 0.01 & 0.35 \\
 4 & 8224 &183576 & 0.015 & 1:1 & 0.40 & 0.78 & 0.18 & 0.61 & 0.01 & 0.27 \\
\end{tabular}
\end{table*}

\end{document}